\documentclass[reprint,amsmath,amssymb,aps,pre,showpacs,twocolumn,floatfix]{revtex4}
\usepackage{epsfig}
\usepackage{graphicx}
\usepackage{makeidx}
\usepackage{chemarr}
\usepackage{dcolumn}
\usepackage{bm,multirow}
\usepackage{color}
\usepackage{bm}

\def\prl#1#2#3{{ Phys. Rev. Lett.} {\bf #1}, #2 (#3)}

\def\pre#1#2#3{Phys. Rev. E {\bf #1}, #2 (#3)}

\def\pnas#1#2#3{Proc. Natl. Acad. Sci. U.S.A {\bf #1}, #2 (#3)}

\def\physd#1#2#3{Physica D {\bf #1}, #2 (#3)}

\def\ch#1#2#3{Chaos {\bf #1}, #2 (#3)}
\def\nat#1#2#3{Nature {\bf #1}, #2 (#3)}
\def\natphys#1#2#3{Nature Phys. {\bf #1}, #2 (#3)}

\def\sci#1#2#3{{ Science} {\bf #1}, #2 (#3)}
\def\neurosci#1#2#3{{ J. Neurosci.} {\bf #1}, #2 (#3)}
\def\plos1#1#2#3#4{{ PloS One} {\bf #1}{(#2)}, #3 {(#4)}}
\def\epjb#1#2#3{Eur. Phys. J. B {\bf #1}, #2 (#3)}
\def\neurophysio#1#2#3{{ J. Neurophysiol.} {\bf #1}, #2 (#3)}

\def\Erdos{Erd\H{o}s}

\def\epsilon{\varepsilon}

\def\beqr{\begin{eqnarray}}
\def\eqnr{\end{eqnarray}}
\def\beq{\begin{equation}}
\def\bc{\begin{center}}
\def\ec{\end{center}}
\def\eqn{\end{equation}}
\begin{document}
\title{Scaling behaviour in probabilistic neuronal cellular automata}
\author{Kaustubh Manchanda$^{1}$, Avinash Chand Yadav$^{1}$ and Ramakrishna Ramaswamy$^1$$^,$$^2$}
\affiliation{$^{1}$School of Physical Sciences, Jawaharlal Nehru University, New Delhi 110067, India\\
$^2$University of Hyderabad, Hyderabad 500 046, India.}
%\date{\today}

\begin{abstract}
We study a neural network model of interacting stochastic discrete two--state cellular automata on a regular lattice. The system is externally tuned to a critical point which varies with the degree of stochasticity (or the effective temperature). There are avalanches of neuronal activity, namely spatially and temporally contiguous sites of activity; a detailed numerical study of these activity avalanches is presented, and single, joint and marginal probability distributions are computed.  At the critical point, we find that the scaling exponents for the variables are in good agreement with a mean--field theory. 
\end{abstract}

\maketitle

\section{Introduction}
Networks in a variety of natural settings
are currently the subject of intense enquiry and span a variety of disciplines, 
ranging from biology \cite{yang:chaos, barabasi:nature, wang:physd}, ecology \cite{jain:pnas}, engineering \cite{peter, albert, walter} to social  interactions \cite{park:pre}. 
The manner in which such networks arise, given the diversity of settings in which these are observed, is an additional issue. The idea that such structures are in some sense inevitable once a set of simple rules and interactions is specified, namely that there is ``self--organization'' at work has been a theme that has also been studied in diverse settings \cite{bak:book}. 

The neural network structure---the nature and significance of the interactions and connections in neuronal systems---has long been recognized and has been an area of study for several decades. There has been extensive work on the dynamical properties of neural networks, in
particular those that emerge as a result of evolving connection strengths \cite{levina:nature, geisel:prl}, the response to external stimuli \cite{kincopel:nature}, and the interplay between topology and intrinsic
nodal dynamics \cite{k1, k2}. A specific question is on how the response of a sensory system to external stimuli is optimized \cite{kincopel:nature, stevens:book, buckley:prl, gollo_rapid}, particularly in the context of information processing \cite{newman:sci} and whether the self--organized network is essential to this feature. When a neural network is in a critical state, the avalanches of spiking activity \cite{gerald:jneuro} are critical, namely they are characterized by
power--law distributions, and the role of network topology \cite{restrepo:prl} appears to be limited. Studies of avalanche dynamics have been both theoretical \cite{novikov:pre,  freeman:neuroscimeth, meisel:ploscompbio, friedman:prl} and experimental: for instance, it has been suggested that the propagation of spontaneous activity recorded from the rat cortex can be described in terms of critical avalanches with an event--size exponent of 3/2 \cite{beggs:neurosci, plenz:neurosci}. 

How does the global dynamics of such networks change when the nature of the interaction is altered? This question is explored here through the use of both analytic and numerical tools and is the main focus of the present work. Similar issues have been examined in other recent works \cite{gollo:plos, gollo:pre} although the geometry of the model and the dynamics differ significantly from the present case.  In the following section, we describe the system studied here. This builds upon the model introduced earlier by Kinouchi and Copelli (KC) \cite{kincopel:nature} by introducing stochasticity in the deexcitation process: excited neurons return to the silent state with some probability, reflecting the inherently stochastic nature of cellular and subcellular dynamics. In concordance with earlier work  \cite{kincopel:nature}  there is a non-equilibrium phase transition, and here  we determine the critical dynamics by computing the Lyapunov exponent. These results are presented in Section III, and this is followed by a study of the avalanches of neuronal activity in Section IV. A mean--field theory for the stochastic transition and a general methodology relating the scaling exponents are also given. The final Section V summarizes our results.

\section{The generalized KC model}
The neuronal cellular automaton model introduced by Kinouchi and Copelli is a quenched \Erdos--R\'enyi network of excitable elements \cite{kincopel:nature}. Nodes in the network can be in one of $m$ states: 0 (silent), 1 (excited or active) or $m -2$ refractory states (which are denoted $-( m -2),  \dots, -1$ for convenience). The dynamics results from the following three processes:  
\begin {enumerate}
\item External driving or noise as a stimulus can cause a silent node to become active. This 0$\to$1 transition occurs at each time step with probability $\eta=1-\exp(-r)$, where $r$ is the stimulus rate.

\item At each time step, a silent node $i$ is excited (namely 0$\to$1) if it receives a stimulus from one of its active neighbors $j$. This can occur with probability $A_{ij}$, the weight of the connection.  
  
\item Spontaneous transitions that can be of two kinds, namely, an excited state becomes refractory ( 1$\to -(m-2)$ ) with unit probability and further, a node in state $l$ moves to $l+1$ if $-(m-2)\le l \le -1$, also with unit probability.

\end {enumerate}

Network response $F$ is calculated by the density of active sites $\rho_{t}$ averaged over a time window $T$. Measure of fluctuations in $F$ is the susceptibility $\chi=\langle F^2\rangle-\langle F\rangle^2$, while the spontaneous activity at $\eta$ = 0 is denoted by $F_{0}$. This nonequilibrium system undergoes a continuous phase transition from a moving phase (corresponding to sustained neuronal activity) to an absorbing phase (where all nodes are in silent state) as a function of the average branching ratio \cite{kincopel:nature}. For every node in the network, the branching ratio gives the average number of excitations created by that element at the subsequent time step. Similar interactions have been used to model epidemics on contact networks \cite{may:book}. 

The relationship between criticality and network structure is crucial, since the response of this network to external stimuli---and hence the dynamical range---is optimal when the spontaneous activity of the network is critical \cite{kincopel:nature}. However, in a generalization of the KC model that accounts for the integration of excitatory inputs \cite{gollo_rapid}, the order of the phase transition itself appears to change: the transition becomes discontinuous and results in bistability in the system, with a history--dependent dynamical range apparently being optimized in this bistable regime. The largest eigenvalue $\lambda$ of the network connectivity matrix is the ``universal'' control parameter \cite{restrepo:prl}. Furthermore, if the connectivity matrix is symmetric with nonzero elements all equal (namely all connections have equal weights) then $\lambda$ is proportional to the average degree of the network \cite{ott:pre}. 

We study a variation on the KC model on a two--dimensional square lattice using both the nearest neighbor (namely Von Neumann) as well as the Moore connectivity. For simplicity we take $m$ = 2; nodes can be either silent or excited. Further, the external stimulus is  absent, namely $\eta$ = 0. Based on the electrophysiological differences, neurons can be broadly categorised as bursting or spiking. Considering a neuron as a discrete dynamical unit, excited state for a single time step characterizes the spiking behaviour whereas multiple consecutive spikes of a bursting neuron preceding a refractory period are mimicked by a prolonged active state lasting several time steps \cite{k1, k2}. To account for both these behaviours in the present network, the transition from the active to the silent state (cf. process 3 in the above dynamics) is considered as a stochastic process. This transition is taken to occur with probability $\mu \in [0, 1]$, $\mu$ = 0 being the ``zero--temperature'' limit, while $\mu = $ 1 corresponding to the case studied earlier by KC for a spiking neuron. The elements of the symmetric connectivity matrix, the $A_{ij}$'s, are drawn from an uniform distribution in $[0,1]$, and we rescale the largest eigenvalue $\lambda$ of this matrix appropriately to use as a control parameter \cite{restrepo:prl}. This rescaling is done in a way that the largest eigenvalue of the new adjacency matrix lies between 0 and 2 with $\lambda$ = 1 being the critical point. The elements of this matrix are kept fixed throughout the simulation implying quenched disorder. We choose to use $\lambda$ as the control parameter to further extend our work to the study of real networks with heterogeneous degree distributions where the branching ratio fails to predict criticality \cite{copelli:epjb}. Also, with the use of $\lambda$ as the control parameter, our calculations become independent of the system size. 

\begin{figure*}[t]
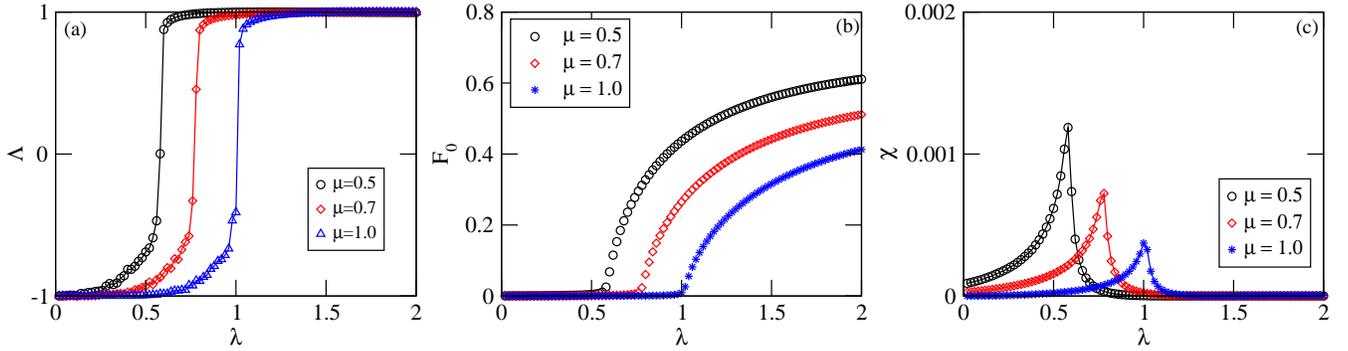

  \centering
  \scalebox{0.5}{\includegraphics{fig1a.eps}}
  \scalebox{0.5}{\includegraphics{fig1b.eps}}
  \scalebox{0.5}{\includegraphics{fig1c.eps}}
   \caption{(a) Boolean Lyapunov exponent $\Lambda$ for $\mathcal{T}_{le}$ = $10^{4}$ and $n$ = $10^{3}$, (b) Spontaneous activity $F_{0}$ averaged for $10^3$ time steps and (c) Susceptibility $\chi$ with control parameter $\lambda$ for different values of $\mu$ for $10^6$ nodes, $\eta$ = 0 and Moore neighborhood.}
\label{fig1}
\end{figure*}

\section{Results}
In the limit $\mu \to$ 1, the present model reduces to that studied by KC. In order to further characterize the critical dynamics we compute the Boolean Lyapunov exponent \cite{beggs:prl} $\Lambda$ by coevolving two initial network configurations and examining the average variation of the Hamming distance as a function of time. The Boolean Lyapunov exponent is given as
\beq
\Lambda = \frac{1}{\mathcal{T}_{le}}\sum_{i=1}^{n}\log_{2}\left [\frac{d_{out}}{d_{in}}\right],
\label{lyap}
\eqn   
where $n$ is the number of sample (random) perturbations considered, $\mathcal{T}_{le}$ is the number of time steps, and $d_{in}$ and $d_{out}$ are the initial and final Hamming distances. Sufficiently large $\mathcal{T}_{le}$ and $n$ need to be taken for reliable results. Other definitions of the Boolean Lyapunov exponents have also been extensively used earlier to characterise the dynamics of cellular--automata models \cite{franco:pre, bagnoli:modphysc}.

Our results are consistent with a critical point at $\lambda$ = 1 for $\mu$ = 1;  nearby trajectories converged for sub--critical networks, namely $\Lambda<0$ for $\lambda<1$ while the dynamics is chaotic in the super--critical networks: $\Lambda >0$ for $\lambda >1$. Shown in Fig.~\ref{fig1}(a) are the numerical results at this value of $\mu$ showing that both $\Lambda$ and $\lambda$-1 cross the axis at the same point. 

For $\mu \ne$ 1, however, we find that the critical point is no longer at $\lambda$ = 1 and is instead shifted by an amount proportional to $\mu$. The change in the sign of the Boolean Lyapunov exponent, the phase transition in $F_{0}$, and the optimization of the susceptibility shown in Fig.~\ref{fig1}(a)--(c) respectively; all three indicators clearly show a shift in the critical point for different values of $\mu$.  Similar results are obtained for $m > 2$; wherein a node in state 1 (excited) goes to state $-(m-2)$ (refractory) with a probability $\mu$ and its subsequent updates till it reaches the state 0 (silent) are deterministic; an extended period of refractoriness does not affect the critical point \cite{larremore:chaos} in the present setting. In 1D (namely $K$ = 2), the system does not show a phase transition at $\mu$ = 1 for a 3 state model ($m$ = 3) but for $\mu < 1$, the phase transition is recovered \cite{assis:pre}.
\begin{figure}[t]
  \centering
  \scalebox{0.65}{\includegraphics{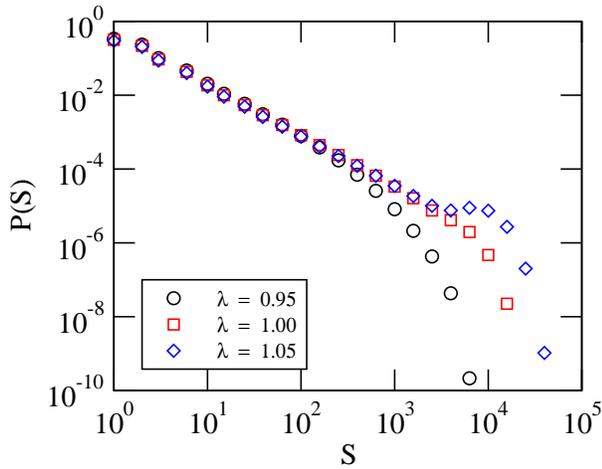}}
  \caption{Plot of avalanche size distribution $P(S)$ for different $\lambda$ for $10^7$ avalanche events.}
\label{fig2}
\end{figure}

\begin{figure}[t]
  \centering
   \scalebox{0.8}{\includegraphics{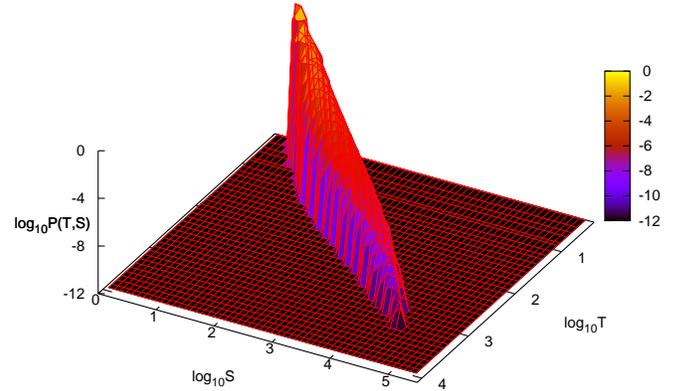}}
  \caption{Joint size--duration probability distribution $P(T,S)$ for total nodes, $N$ = $1024^2$ using log binned data.}
\label{fig3}
\end{figure}

\begin{figure}[t]
  \centering
  \scalebox{0.65}{\includegraphics{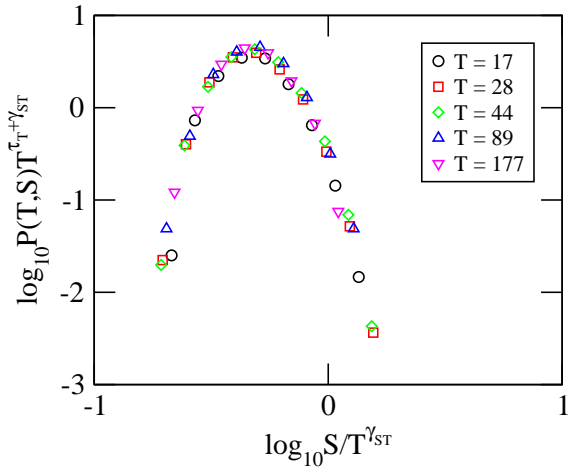}}
  \caption{The data collapse of $P_{T}(S)$ curves for the joint size-duration distribution. Each $P_{T_{0}}(S)$ curve is the intersection of the joint size--duration distribution $P(T,S)$ with the $T = T_{0}$ plane. The collapse width is $w$ = 4.81$ \times 10^{-2}$.}
\label{fig4}
\end{figure}

We confirm the linear variation of the critical value, namely $\lambda_{c} \sim \mu$ both numerically and analytically using mean--field analysis. In the KC model the transition from the silent to the active state (namely 0 $\to$ 1) can take place in two ways: (i) through an external stimulus, and (ii) via active neighbor interactions. On a two dimensional square lattice, the degree distribution is $P(k) = \delta(k-K)$, where $K$ denotes degree of each node. The probability of causing a silent node to become active at any time $t$ via the input it receives from at least one of its active neighbors is given by $p_{t} = 1-[1-(\lambda/K)F_{t}]^K$ where $F_{t}$ denotes the fraction of active nodes at time $t$ and in the mean field approximation $\langle A_{i,j} \rangle$ is $\lambda /K$.

The transition from state 1 $\to$ 0 occurs with probability $\mu$. If $P_{1}(t)$ denotes the configuration probability of active nodes and $P_{0}(t)$ that of silent nodes, the corresponding master equation is
\beq
P_{1}(t+1) = \eta P_{0}(t) + (1-\eta)p_{t}P_{0}(t) + (1-\mu)P_{1}(t).
\eqn
Denoting $P_{1}(t) = F_{t}$, one has $P_{0}(t) = 1-F_{t}$ and in terms of $F_{t}$, the master equation becomes
\beq
F_{t+1} = [1-F_{t}][\eta + (1-\eta)p_{t}] + (1-\mu)F_{t}.
\eqn
In the stationary state $F_{t+1} = F_{t} = F$, giving
\beq
\label{eq1}
\mu F = (1-F)\left[1-(1-\eta)\left(1-\frac{\lambda}{K}F\right)^K\right].
\eqn

In the absence of an external stimulus namely $\eta$ = 0, by setting $\lambda$ as $\lambda_{c} + \epsilon$, in the limit $\epsilon \to 0$, $F \to 0$ we get
\beq
\lambda_{c} \to \mu,
\eqn
and
\beq
F(\lambda) \approx \frac{\lambda - \mu}{C},
\eqn
where the constant $C = \mu + \frac{K-1}{2K}\mu^ 2$. Therefore, since $F(\lambda)$ scales as $|\lambda-\lambda_{c}|^{\beta}$, we get $\beta$ = 1, although the numerical results ($\beta_{num}$ = 0.499 $\pm$ 0.002) are not in good agreement with this estimate. 

At the critical point $\lambda_{c} = \mu$ and with  external stimulus $\eta$, the resulting behaviour is 
\beq
F(r) \approx \sqrt{r/C}
\label{fr}
\eqn
where $C$ is an arbitrary constant. The numerical results are in better agreement with this behaviour, namely the scaling $F(r) \sim r^{1/\delta}$ with $\delta = 2 \approx \delta_{num}$ (not shown here). In the limit $\lambda -\mu \to \epsilon$, near the critical point, $F \to 0$, we find that the susceptibility varies as 
\beq
\chi \sim \frac{1}{|\epsilon|},
\eqn \noindent
namely the scaling $\chi \sim |\epsilon|^{-\gamma}$ ($\epsilon >$ 0) and $\chi \sim |\epsilon|^{-\gamma'}$ ($\epsilon <$ 0) with $\gamma = \gamma'$ = 1. Numerically, however,  we find  $\gamma_{num}$ = 0.89 $\pm$ 0.04, $\gamma_{num}'$ = 1.15 $\pm$ 0.04.

A pertinent observable that has been introduced for such excitable systems is the dynamic range
\beq
\Delta = 10\log_{10}\left[\frac{r_{0.9}}{r_{0.1}}\right],
\label{dr}
\eqn
where the external stimulus $r_x$ leads to response $F_x$.
$\Delta$ measures the range of stimuli that can be discerned by the changes in the network response $F$. Previous studies have reported that the dynamic range attains its maximum value $\Delta^{max}$ at the critical point of the system \cite{kincopel:nature, restrepo:prl}. With the addition of stochasticity, $\Delta^{max}$ occurs at the shifted critical point consistent with Fig.~\ref{fig1}, however, its magnitude appears to be independent of $\mu$  (cf. Eqs.~(\ref{fr}) and (\ref{dr})).

The mean--field exponents obtained so far are $\beta$ = 1, $\gamma$ = 1, and $\delta$ = 2. Now, we consider the cluster size probability distribution which obeys $P(S)\sim S^{-\tau+1} \sim S^{-\tau_{S}}$ (for details refer section \ref{App1}). Below the critical point, the cutoff cluster size is related to correlation length  $\xi \sim |\epsilon|^{-\nu}$ as $S_{co} \sim |\epsilon|^{-1/\sigma} \sim \xi^{D}$, where $\epsilon$ is the control parameter and $D$ is the fractal dimension of cluster. The order parameter $F$ is characterized by 
\beq
\int dS P(S) \sim |\epsilon|^{\beta},
\eqn
where $P(S) \sim S^{-\tau+1}f(S|\epsilon|^{1/\sigma})$ and $f$ a general scaling function. Changing variables to $y=S|\epsilon|^{1/\sigma}$, the above integral reduces to 
\beq |\epsilon|^{(\tau-2)/\sigma} \int dy y^{-\tau+1}f(y),
\eqn
and hence on comparison with $F \sim |\epsilon|^{\beta}$, we infer that $\beta=(\tau-2)/\sigma$.
The different moments of $P(S)$ connect the various exponents via simple algebraic equations,
and with a little computation, it can be seen that  $\alpha=3-2/\sigma$, $\beta=(\tau-2)/\sigma$, $\gamma=(3-\tau)/\sigma$, $D=1/\sigma\nu$, and $\tau=2+1/\delta$ \cite{stauffer, christensen}.
Using the known values of $\delta$ and $\gamma$, it is a simple matter to get
$\tau$ = 5/2 , $\sigma$ = 1/2, and $\alpha$ = -1.

If we assume that the avalanche dynamics on the network is diffusion like \cite{tang}, then since the size of the largest cluster is limited by the linear system size $L$, any perturbation at the critical point takes an average $L^{2}$ steps to diffuse away. So, from the stationarity condition we get, $\gamma/\nu$ = 2, resulting in $\nu$ = 1/2 \cite{tang} and $D$ = 4. Using the hyperscaling relation $D=d-\beta/\nu$ we get $d$ = 6 \cite{univers} as the upper critical dimension.

\section{\label{App1} Avalanche Properties}
Starting with a null configuration (all nodes are in state 0) we select a site at random and make it  active. This random flipping induces activity in the neighborhood which may then propagate, but this will eventually die out so that the system returns to the null configuration.  This entire event constitutes an avalanche and the total activity at each time step is the avalanche signal $s(t)$. If $T$ is the total time, then the total activity during this event
\beq
S=\sum_{t=0}^{T} s(t),
\eqn
 is the avalanche size $S$. We also compute other related  quantities, the area $A$ which is equal to the total number of distinct sites which become active during the avalanche, the energy 
\beq 
E = \sum_{t=0}^{T}s^2(t),
\eqn
and the magnitude
\beq
M= \max{\vert s(t)\vert : 0\le t \le T}.
\eqn

The probability distributions of each of these quantities have been computed; shown in Fig.~\ref{fig2} is the avalanche size distribution $P(S)$ for different values of $\lambda$ where the power--law behaviour in the super--critical, critical and sub--critical regimes can be clearly identified. The scaling relations for these distributions can be defined in general as $P(X) \sim X^{-\tau_{X}}$, where $X$ can be size $S$, duration $T$, area $A$, magnitude $M$ or energy $E$. 

The various exponents are, of course, related. If $X,Y$ are a pair of the above variables and if $\langle X \rangle \sim \langle Y\rangle^{\gamma_{XY}}$, where $\langle \cdot \rangle$ denotes the average value, then there is a simple relation among the exponents $\tau_{X}$, $\tau_{Y}$ and $\gamma_{XY}$, 
\beq
\tau_{X} = 1 - \frac{1-\tau_{Y}}{\gamma_{XY}}.
\label{tx1}
\eqn
Further, using the relationship $\langle Y \rangle \sim X^{\gamma_{YX}}$, one gets
\beq
\tau_{X} = 1 - (1-\tau_{Y})\gamma_{YX},
\label{tx2}
\eqn
and from Eq.~(\ref{tx1}) and (\ref{tx2}), 
\beq
\gamma_{XY} = \frac{1}{\gamma_{YX}}.
\label{g1}
\eqn

The scaling form for the joint probability distribution, $P(X,Y)$, can be obtained via 
\beq
P(\lambda^{a_{X}}X,\lambda^{a_{Y}}Y) = \lambda P(X,Y). 
\label{pxy}
\eqn
where, $a_{X}$ and $a_{Y}$ are the corresponding exponents associated with the arguments of a generalized homogeneous function.

By putting $\lambda \sim X^{-1/a_{X}}$ in Eq.~(\ref{pxy}), we get
\beq
P(X,Y) \sim X^{\frac{1}{a_{X}}}P(Y/X^{a_{Y}/a_{X}}).
\label{pxty}
\eqn  

Comparing this to the expression
\beq
P(X,Y) \sim X^{-(\tau_{X} + \gamma_{YX})}P(Y/X^{\gamma_{YX}}),
\label{pxep}
\eqn
one can see that 
\beq
\gamma_{YX} = \frac{a_{Y}}{a_{X}},~~~ \frac{1}{a_{X}} = -(\tau_{X}+\gamma_{YX}).
\label{gyx}
\eqn

The replacment $X \to Y$ and $Y \to X$ in Eq.~(\ref{gyx}) similarly gives
\beq
\gamma_{XY} = \frac{a_{X}}{a_{Y}},~~~ \frac{1}{a_{Y}} = -(\tau_{Y}+\gamma_{XY}).
\label{gxy}
\eqn
Clearly, Eqs.~(\ref{gyx}) and (\ref{gxy}) can be used to obtain Eqs.~(\ref{tx1}-\ref{g1}). 

We study networks with degree $K$ = 4 and system size $N$ = $1024^2$ at the critical point (for $\mu$ = 1) and compute joint distributions \cite{knezevic:pre,knezevic:prl} for four pairs of the above related quantities, size--duration (see Fig.~\ref{fig3}) as well as size--magnitude, size--area  and size--energy (which are not shown here). For each of these joint probability distributions $P(X,Y)$ we fix one variable $Y$ to obtain the marginal-$X$ distribution, $P_{Y}(X)$. This obeys the scaling relation Eq.~(\ref{pxep}).

A typical data collapse of the marginal-$X$ distributions is shown in Fig.~\ref{fig4}, and critical exponents obtained through this analysis are given in Table \ref{tab1}. As can be seen, these are in good agreement with the corresponding exponents obtained through the scaling relations (within statistical error). The collapsing width for the various curves was also calculated \cite{knezevic:pre}. Although in the present study the network is externally driven to criticality by varying the control parameter $\lambda$, avalanches showing power--law behaviour also form an integral part of the study of self--organized critical neuronal systems \cite{levina:nature}.

\begin{table}[t]
\centering
\begin{tabular}{ccc}
\hline 
\hline
  Exponent & Measured value & Scaling relation\\
\hline 
  $\tau_{S}$ & $1.49 \pm 0.03$ & ---\\ 
  $\tau_{T}$ & $1.82 \pm 0.06$ & 1.77\\
  $\tau_{A}$ & $1.69 \pm 0.06$ & 1.72\\
  $\tau_{M}$ & $1.97 \pm 0.07$ & 2.10\\
  $\tau_{E}$ & $1.35 \pm 0.02$ & 1.35\\ 
  $\gamma_{ST}$ & $1.527 \pm 0.002$ & --- \\
  $\gamma_{SA}$ & $1.474 \pm 0.003$ & --- \\
  $\gamma_{SM}$ & $2.25 \pm 0.03$ & ---\\
  $\gamma_{SE}$ & $0.7712 \pm 0.0005$ & ---\\
\hline
\hline
\end{tabular}
 \caption{Exponents characterizing avalanche properties. Measured values are obtained from data collapse while the values mentioned in the third column are obtained from the scaling relation Eq.~(\ref{tx1}).}
\label{tab1}
\end{table}

\section{Discussion and Summary}
Recent efforts to understand the operation of neuronal systems have examined a number of models wherein the feature of self--organization is important. This would be one way of understanding the emergence of complexity in the context of neuroscience. Although automaton models are by their very nature caricatures of how a biological neuron operates, they capture some of the essential features. Furthermore, the manner in which neuronal communication operates can also be modeled within the framework of coupled automata.

Here we investigated the automaton model introduced by Kinouchi and Copelli \cite{kincopel:nature}. While it is known that the critical point of this model is invariant under the change of topology  \cite{gerald:jneuro},  we alter the nodal dynamics of the KC model and consider its impact on criticality.  Other recent studies have also discussed the effect of these variations in a different context \cite{gollo:pre}, underscoring the considerable current interest in such problems.  We incorporate additional stochasticity in the deexcitation transition from the active to the silent state and the primary  effect of this is to shift the point of criticality without destroying it. This extended model thus retains the main features of criticality and emergence. Although the effect of having a variable number of refractory states on the critical point in the KC model has been investigated \cite{larremore:chaos}, in the present case, the dynamics tends to hold the neuron in the excited state for a longer duration thus taking into account the bursting nature as well. Furthermore, owing to the inherent stochasticity in the intrinsic dynamics of a neuron, the duration of bursts may also differ (\cite{burst:var} and references therein). Indeed, experiments suggest that there exists variability in the burst durations \cite{new}. 

Activity in the network is characterized through avalanches, namely sets of firing nodes that are connected spatially and temporally. Probability distributions for a variety of quantities that quantify the avalanches can be described through a set of exponents which we have computed in a two dimensional realization of this model. We also developed a mean--field model for the dynamics, and show that at the critical point the numerical results are in good agreement with the theoretical predictions, and further, with the exponents obtained for branching processes \cite{harris:book, stanley:prl}. The KC model established a theoretical framework for the optimization of dynamical range at criticality, and this has been subsequently verified experimentally \cite{expt1, expt2}. It would also be of interest to see if optimization of susceptibility can also be demonstrated experimentally in such models.  \\

\begin{acknowledgments}KM and ACY acknowledge the support from UGC, India and CSIR, India respectively through the award of Senior Research Fellowship. We thank Gaurav Shrivastav and Rupesh Kumar for valuable discussions. Our presentation, and indeed the work itself has benefited greatly from the comments of anonymous referees.
\end{acknowledgments}


\begin{thebibliography}{99}
\bibitem{univers} The final set of exponents reads as: $\alpha$ = -1, $\beta$ = 1, $\gamma$ = 1, $\delta$ = 2, $\nu$ = 1/2, $\tau$ = 5/2, $\sigma$ = 1/2, and $D$ = 4 which fall into the universality class of percolation models on Bethe lattice \cite{stauffer, christensen}.
\end{thebibliography}
\end{document}